\documentclass{elsart}
\usepackage{graphicx,amssymb,amsfonts} \journal{Physica A}

\begin{document}

\begin{frontmatter}
\title{A Stabilization Theorem for Dynamics of Continuous Opinions}

\author{Jan Lorenz}
\address{University of Bremen, Department of Mathematics, Germany,
\texttt{math@janlo.de}}
\thanks{The author thanks the Friedrich-Ebert-Stiftung (Bonn, Germany) for
a grant.}

\begin{abstract}
A stabilization theorem for processes of opinion dynamics is
presented. The theorem is applicable to a wide class of models of
continuous opinion dynamics based on averaging (like the models of
Hegselmann-Krause and Weisbuch-Deffuant). The analysis detects
self-confidence as a driving force of stabilization.
\end{abstract}


\end{frontmatter}

\newcommand{\eps}{\varepsilon}
\newcommand{\conv}{\mathrm{conv}}
\newcommand{\diam}{\mathrm{diam}}
\newcommand{\dasd}{\mathrm{d}}
\newcommand{\ran}{\mathrm{range}}
\newcommand{\wei}{\mathrm{weight}}
\newcommand{\ci}{\mathcal{I}}
\newcommand{\cj}{\mathcal{J}}
\newcommand{\n}{\underline{n}}
\newcommand{\m}{\underline{m}}
\newcommand{\p}{\underline{p}}

\section{Modelling of opinion dynamics}
Consider a group of $n$ agents each having an opinion about a
certain issue. The agents may revise their opinions according to
the opinions of other agents. If revising goes on we have a
process of opinion formation. The understanding of phenomena like
stabilization of opinions distribution, finding a consensus,
polarization into opinion clusters, extremism or spreading of
minority opinions is of interest in sociology, political science
and economics (e.g. price setting or customer's opinions about
brands). Mathematical models and their analysis should detect
driving forces of opinion dynamics.

Here, we consider the continuous opinion approach where the
opinion space is a real interval, see
\cite{JASA:degroot74,JAP:chatterjee77,jasss:krauheg02,cond:redner02,WeisbuchEtal2002}.
Thus, the opinion dynamic can be driven by compromising.

For the model of continuous opinion dynamics we consider
$\underline{n}:=\{1,\dots,n\}$ agents who discuss their opinions.
We call $X(t) \in \Rset^{n}$ an {\em opinion profile \/} at time
step $t\in\Nset_0$, where $X_i(t)$ represents the opinion of agent
$i$.

\begin{defn}[confidence matrix]
Let $X(t)$ be an opinion profile at time step $t\in\Nset_0$. A
matrix $A(X(t),t)\in\Rset_{\geq 0}^{n\times n}$ is called {\em
confidence matrix\/} if it is row-stochastic.
\end{defn}

The entry $A(X(t),t)_{[i,j]}$ represents the weight (or
confidence) that agent $i$ distributes to the opinion of agent
$j$. Notice that the confidence matrix is a function of the actual
opinion profile and of the specific time step. Let $X(0) \in
\Rset^{n}$ be a starting opinion profile. The {\em process of
continuous opinion dynamics} is the series of opinion profiles
$(X(t))_{t\in\Nset_0}$ recursively defined through
\[ X(t+1) = A(X(t),t)X(t). \] Thus, each new opinion is a weighted
arithmetic mean of all the old opinions. It holds $X(t+1) =
A(X(t),t) \cdots A(X(0),0) X(0)$ by iteration.

This very general agent-based setting gets explicit by defining
how the confidence matrix is constructed. The setting also
contains models with heterogeneous agents, underlying network
structures and various updating rules, as long as repeated
averaging drives the dynamic. Further on $m$-dimensional opinions
can be modelled by regarding $X(t) \in \Rset^{n\times m}$.

DeGroot \cite{JASA:degroot74} analyzes the model for fixed $A$ and
gives conditions for reaching consensus. Chatterjee and Seneta
\cite{JAP:chatterjee77} derived some generalizations for $A(t)$ in
the sense of hardening of positions.

In this paper we want to treat the much more complicated profile
dependent case, where no analytical results are available. We will
point out weak but sufficient conditions on the confidence
matrices such that the process converges to a fixed opinion
profile. But this conditions are not necessary. These conditions
are for all $t \in \Nset_0$

\begin{enumerate}

\item \textbf{Every agent got a little bit of self-confidence.}
The diagonal of $A(X(t),t)$ is positive. For every agent
$i\in\underline{n}$ it holds $a_{ii} > 0$.  \label{prop:posdia}

\item \textbf{Confidence is mutual.} Zero-entries in $A(X(t),t)$
are symmetric. For every two agents $i,j\in\underline{n}$ it holds
$a_{ij} > 0 \Leftrightarrow a_{ji}
> 0$. \label{prop:syminc}

\item \textbf{Positive weights do not converge to zero.} There is
$\delta>0$ such that the lowest positive entry of $A(X(t),t)$ is
greater than $\delta$. \label{prop:posmin}

\end{enumerate}

In the bounded confidence model of Hegselmann-Krause
\cite{jasss:krauheg02} the confidence matrix is defined for $\eps
>0$ and an opinion profile $X\in\Rset^n$ as
\[
 A(X)_{ij} :=
 \left\{ \begin{array}{cl}
   \frac{1}{|I(i,X)|} \quad & \textrm{if } j\in I(i,X) := \{j \in \n \,|\, |X_{i} - X_{j}| \leq \eps \}   \\
   0 & \textrm{otherwise}
\end{array} \right. \\
\] In the basic model of Weisbuch, Deffuant et al.
\cite{WeisbuchEtal2002} two randomly chosen agents $i,j\in\n$
interact in each time step. They adjust their opinions if
\mbox{$|X(t)_i - X(t)_j| \leq \eps$} by a step of $\mu|X(t)_i -
X(t)_j|$ towards each other ($0 < \mu < 0.5$). Thus a confidence
matrix in one time step is the unit matrix besides the entries
$a_{ii}=a_{jj}=1-\mu$ and $a_{ij}=a_{ji}=\mu$.

Thus, it is easy to check that both the Hegselmann-Krause and the
basic Weisbuch-Deffuant model fulfill conditions
(\ref{prop:posdia})-(\ref{prop:posmin}).

\section{The Stabilization Theorem}

For abbreviation, we define for a series of matrices
$(A(t))_{t\in\Nset_0}$ the {\em accumulation\/} from time step
$t_0$ to $t_1$ as $A(t_0,t_1) := A(t_1-1) A(t_1-2) \cdots A(t_0+1)
A(t_0)$. A {\em consensus matrix} should be a row-stochastic
matrix with equal rows. With definition $A(t) := A(X(t),t)$ we can
write $X(t)=A(0,t) X(0)$. We will show that $\lim_{t\to {\infty}}
A(0,t)$ converges to a constant matrix. This implies that
$(X(t))_{t\in\Nset_0}$ converges to a constant opinion profile.

\begin{thm}\label{thm}
Let $(A(t))_{t\in\Nset_0} \in \Rset_{\geq 0}^{n\times n}$ be a
series of confidence matrices. If each matrix fulfills properties
(\ref{prop:posdia})-(\ref{prop:posmin}), there exists a time step
$t_0$ and pairwise disjoint classes of agents \mbox{$\ci_1 \cup
\dots \cup\ci_p =\underline{n}$} such that
\[
\lim_{t\to{\infty}} A(0,t) = \left[ \begin{array}{ccc}
              K_1 &        & 0 \\
                  & \ddots &  \\
0                 &        & K_p \\
\end{array} \right] A(0,t_0),
\]
and $K_1,\dots,K_p$ are quadratic consensus matrices in the sizes
of $\ci_1,\dots,\ci_p$. (For the block structure we must sort
matrix indices according to $\ci_1,\dots,\ci_p$.)
\end{thm}

In front of the proof some explanations and necessary
propositions: If we multiply the consensus matrix $K_i$ with an
arbitrary vector then we get a vector with all entries equal. Thus
the theorem says that every starting opinion profile develops to a
time step $t_0$, where the agents split into some independent
classes. The opinions of the agents in these classes converge to
consensus.

For a matrix $A\in\Rset_{\geq 0}^{n\times n}$ we say that two
agents $i,j\in\underline{n}$ {\em communicate}, if there exist
agents $i=i_1,\dots,i_k=j \in \underline{n}$ such that for all $l
= 1,\dots,k-1$ the agents $i_{l}$ and $i_l+1$ trust each other
($a_{i_{l}i_{l+1}}>0$). It is easy to see that the set of agents
$\underline{n}$ divides for every opinion profile into
self-communica\-ting classes $\ci_1,\dots,\ci_p$. This means, each
agent communicates with every other agent in his class, but with
no agent outside. Notice that the structure of self-communicating
classes of indices depends only on the zero-pattern of the matrix.

We need the following three propositions to prove the theorem.

\begin{prop} \label{posdiag}
Let $(A(t))_{t\in\Nset_0}\in \Rset_{\geq 0}^{n \times n}$ be a
series of matrices fulfilling condition (\ref{prop:posdia}), then
there exists a series of time steps $t_0 < t_1 < t_2 < \cdots$
such that $A(t_0,t_1), A(t_1,t_2), \dots$ got the same
zero-pattern. Let $\ci_1,\dots,\ci_p$ be the self-communicating
classes of agents of the matrices $A(t_0,t_1), A(t_1,t_2), \dots$.
If we sort the agents of every matrix by simultaneous row and
column permutations, then we got a block matrix with strictly
positive blocks on the diagonal
($A(t_k,t_{k+1})_{[\ci_i,\ci_i]}>0$ for all $k\in\Nset_0$,
$i\in\underline{p}$) and zero-blocks at all other positions
($A(t_k,t_{k+1})_{[\ci_i,\ci_j]}=0$ for all $k\in\Nset_0$ and
$i,j\in\underline{p}$, $i\neq j$).
\end{prop}

\begin{prop}\label{symmfolg}
Let $(A(t))_{t\in\Nset_0} \in \Rset_{\geq 0}^{n\times n}$ be a
series of confidence matrices fulfilling conditions
(\ref{prop:posdia})-(\ref{prop:posmin}). Then it holds for every
two time steps $t_0 < t_1$ that the lowest positive entry of
$A(t_0,t_1)$ is greater than $\delta^{n^2-n+2}$.
\end{prop}

\begin{prop} \label{scramfolg}
Let $(A(t))_{t\in\Nset_0} \in \Rset_{\geq 0}^{n\times n}$ be a
series of row-stochastic matrices and let $\delta_t
> 0$ be a series with $\sum_{t=0}^{\infty} \delta_t={+\infty}$. If it
holds for all $t\in\Nset_0$ that $\min_{i,j} \sum_{k=1}^n
\min\{a(t)_{ik},a(t)_{jk}\} \geq \delta_t$ then there exists a
consensus matrix $K$ such that $\lim_{t \to {\infty}} A(0,t) = K$.
\end{prop}

Proofs are in the appendix.

\begin{pf} (of Theorem \ref{thm}) Proposition \ref{posdiag} gives
us time steps $t_0 < t_1 < t_2 < \dots$ and classes of indices
$\ci_1,\dots,\ci_p$ such that each matrix $A(t_i,t_{i+1})$ got
positive blocks $A(t_i,t_{i+1})_{[\ci_j,\ci_j]}$ for all
$j\in\underline{p}$ and zeros elsewhere.

\sloppypar{ From proposition \ref{symmfolg} and condition
(\ref{prop:syminc}) we can derive that the lowest entry of each
$A(t_i,t_{i+1})_{[\ci_j,\ci_j]}$ is greater than
$\delta^{n^2-n+2}$. For the series
$(A(t_i,t_{i+1})_{[\ci_j,\ci_j]})_{i\in\Nset_0}$  the assumptions
of proposition \ref{scramfolg} are fulfilled for all
$j\in\underline{p}$.}

Further on it holds that $[\dots
A(t_1,t_2)A(t_0,t_1)]_{[\ci_j,\ci_j]} = \dots
A(t_1,t_2)_{[\ci_j,\ci_j]}A(t_0,t_1)_{[\ci_j,\ci_j]}$ due to the
block structure. Thus there is a consensus matrix $K_j$ such that
$\lim_{i\to{\infty}}A(t_0,t_i) = K_j$. \qed
\end{pf}

Thus, the convergence to an opinion profile with consensus
subgroups is proved for the model of Hegselmann-Krause, where it
was only proved for the 1-dimensional case with no
generalizations, and for the basic Weisbuch-Deffuant, which was
only observed in simulation. Ben-Naim et al.\ \cite{cond:redner02}
propose a differential equation for the opinion distribution for
the basic Weisbuch-Deffuant model and have other arguments for
stabilization. But they treat idealized ${+\infty}$ agents. We
focus on the dynamic of a finite number of agents using completely
different technics and generalizing to various models.

The theorem more colloquial: A process of continuous opinion
dynamics stabilizes when (\ref{prop:posdia}) each agent has a
little bit of self confidence, (\ref{prop:syminc}) confidence is
mutual and (\ref{prop:posmin}) these two conditions do not fade
away by convergence to zeros. This detects self-confidence as a
driving force of stabilization in continuous opinion dynamics in a
completely analytical way. If we had no self-confidence periodic
behavior may happen. If we drop mutual confidence we might imagine
an open-minded agent between two narrow-minded agents (the
open-minded trusts the narrow-minded but they do not trust him).
The open-minded may hop around in the space between but will not
be converging.

The theorem secures stabilization for simulation of further models
basing on averaging and fulfilling properties
(\ref{prop:posdia})-(\ref{prop:posmin}) which may contain
multidimensional opinions, heterogeneous agents, network
structures and sophisticated updating rules.

\appendix

\section{Appendix}

\subsection{Proof of Proposition \ref{posdiag}}

\sloppypar{Notice that for any two non-negative matrices with
positive diagonals \mbox{$A,B\in\Rset_{\geq 0}^{m\times m}$} it
holds that every entry which is positive in $A$ or in $B$ is also
positive in $AB$. Therefore, more and more positive entries appear
in $A(0,t)$ monotonously increasing with $t$. Thus, once there
will be a time step $t_0^\ast$ in which the maximum number of
positive entries in $A(0,t)$ for all $t\in\Nset$ is reached. And
it is clear that no matrix $A(t)$ with $t\geq t^\ast_0$ got a
positive entry, where $A(0,t^\ast_0)$ has got a zero-entry.}

If we look at the series $(A(t))_{t \geq t^\ast_0}$, we find
another time step $t^\ast_1$, such that $A(t^\ast_0,t^\ast_1)$ has
reached again the maximum number of positive entries, but there
are less or equal positive entries as in $A(0,t^\ast_0)$.

If we continue like this we get a series
$A((t^\ast_i,t^\ast_{i+1}))_{i\in\Nset_0}$ of accumulations in
which positive entries vanish monotonously. Thus, once there will
be a time step $t^\ast_k =: t_0$ for which the minimum of positive
entries is reached and so with $t_i := t^\ast_{i+k}$ we got the
asserted series of time steps.

For proving the block structure, we first notice that it is clear
(due to the definition of self-communicating classes) that
$A(t_k,t_{k+1})_{ij} = 0$ for all $i,j \in \n$ coming from
different self-communicating classes. The last thing to show is,
that for every self-communicating class $\ci_l$ it holds that
$A(t_k,t_{k+1})_{[\ci_l,\ci_l]}$ is strictly positive. For all
$k\in\Nset_0, l\in\p$ the matrix $A(t_k,t_{k+1})_{[\ci_l,\ci_l]}$
is primitive (that means that one power is positive) because all
agents are communicating and the diagonal is positive. The
primitivity property depends only on the zero pattern of a matrix,
which is equal in $A(t_k,t_{k+1})_{[\ci_l,\ci_l]}$ for every
$k\in\Nset$. Thus, there exists $z\in\Nset$ such that
$A(t_0,t_z)_{[\ci_l,\ci_l]}=A(t_{z-1},t_z)_{[\ci_l,\ci_l]}\cdots
A(t_0,t_1)_{[\ci_l,\ci_l]}$ is strictly positive. Thus,
$A(t_k,t_{k+1})_{[\ci_l,\ci_l]}$ must be strictly positive for all
$k$ because otherwise, there were less positive entries than in
later accumulations, which is a contradiction to the minimality of
positive entries proved before. \qed

\subsection{Proof of Proposition \ref{symmfolg}}

Let $t_0 < t_1 $ and $n^{\ast}:=t_1-t_0$. Let $\mu(A)$ be the
lowest positive entry of the non-negative matrix $A$. With
condition (\ref{prop:posmin}) it holds that $\mu(A(t_0,t_1)) \geq
\mu(A(t_1-1)) \cdots \mu(A(t_0)) \geq \delta^{n^{\ast}}$. If
$n^{\ast} \leq n^2 - n +2$ we are ready. Otherwise we will need at
least $n^2 - n +2$ multiplications in $A(t_0,t_1)$ to reach
$\mu(A(t_0,t_1)) < \delta^{n^2 - n +2}$. We will show below that
in each step where the positive minimum sinks we must lose one
zero entry. Thus $\mu(A(t_0,t_1))<\delta^{n^2 - n +2}$ implies
that $A(t_0)$ must have $n^2 - n +2$ zeros more than $A(t_0,t_1)$
and thus can not have a positive diagonal, a contradiction to
condition (\ref{prop:posdia}).

In formal terms we have to show for two confidence matrices $A,B
\in \Rset_{\geq 0}^{n\times n}$ fulfilling conditions
(\ref{prop:posdia}) and (\ref{prop:syminc}) that it holds
\begin{equation}
\mu(AB) < \mu(B) \Longrightarrow \textrm{ $\exists \, (i,j)$ such
that $(AB)_{ij}>0$ and $B_{ij}=0$.} \label{lesszero}
\end{equation}
Due to property (\ref{prop:posdia}) it holds that all non-zero
entries in $B$ are also non-zero entries in $AB$. To prove
(\ref{lesszero}) we assume that the zero patterns of $AB$ and $B$
are equal and derive $\mu(AB) \geq \mu(B)$.

Let $i,j\in\n$ be indices such that $(AB)_{ij}
> 0$ (and $b_{ij}>0$) we can conclude
\begin{eqnarray*}
(AB)_{ij} &=& \sum_{k\in\n \,\textrm{\scriptsize with}\, b_{kj}>0}
a_{ik}b_{kj} \geq (\min_{k\in\n \,\textrm{\scriptsize with}\,
b_{kj}>0} b_{kj})( \sum_{k\in\n \,\textrm{\scriptsize with}\,
b_{kj}>0} a_{ik}) \\
& \stackrel{(\ast)}{=} & \min_{k \,\textrm{\scriptsize with}\,
b_{kj}>0} b_{kj} \geq \mu(B)
\end{eqnarray*}
Equality ($\ast$) holds by $\sum_{k\in\n \,\textrm{\scriptsize
with}\, b_{kj}>0} a_{ik} = 1$ which holds by the following
argument.
\[
b_{kj} = 0 \Rightarrow (AB)_{kj}=0 \Rightarrow \sum_{l=1}^n
a_{kl}b_{lj} = 0 \stackrel{b_{ij}>0}{\Longrightarrow} a_{ki} = 0
\stackrel{(2)}{\Longrightarrow} a_{ki}=0 \qed
\]

\subsection{Proof of Proposition \ref{scramfolg}}

For $A\in\Rset^{n\times n}$ we can define the row-diameter
$\dasd(A)$ as the maximum Euclidean distance of two arbitrary rows
in $A$. It can be shown that multiplication from the left with a
row-stochastic matrix $A\in\Rset^{n\times n}$ to a matrix
$B\in\Rset^{n\times n}$ shrinks the row-diameter of $B$ in this
way

\begin{equation}\label{shrinkinglemma}
\dasd(AB) \leq \left( 1-\min_{i,j} \sum_{k=1}^n
\min\{a_{ik},a_{jk}\}\right) \dasd(B)
\end{equation}

Now we can conclude
\[
\dasd(A(0,t+1)) \leq (1-\delta_t)\dasd(A(0,t)) \leq e^{-\delta_t}
\dasd(A(0,t)) \leq  e^{-\sum_{i=0}^t \delta_t} \dasd(A(0)).
\]
Thus $\lim_{t\to{\infty}}\dasd(A(0,t))=0$ and this leads in our
row-stochastic case to $\lim_{t\to{\infty}} A(0,t) = K$ consensus
matrix. \qed

Equation (\ref{shrinkinglemma}) is a more dimensional version of
the well known shrinking lemma, seen for example in
\cite{CDE:krause00}. For a proof see \cite{dipl} (p. 22-23, Satz
2.4.7).

\end{document}